\documentclass{ws-ijbc}
\usepackage{ws-rotating}
\usepackage{amssymb}    
\begin{document}

\markboth{Gerlach, Eggl, Skokos}{Integration of variational equations}

\title{Efficient integration of the variational equations of
  multi-dimensional Hamiltonian systems: Application to the
  Fermi-Pasta-Ulam lattice}

\author{ENRICO GERLACH}
\address{Lohrmann Observatory, Technical University Dresden, \\D-01062
  Dresden, Germany\\enrico.gerlach@tu-dresden.de}

\author{SIEGFRIED EGGL} \address{Institute for Astronomy,
  University of Vienna, \\
  T\"{u}rkenschanzstr.~17,
  A-1180, Vienna, Austria\\
  siegfried.eggl@univie.ac.at}

\author{CHARALAMPOS SKOKOS} \address{Max Planck Institute for the
  Physics of Complex Systems,\\ N\"{o}thnitzer Str. 38, D-01187,
  Dresden, Germany\\ and\\ Center for Research and Applications of
  Nonlinear Systems,\\ University of Patras, GR-26500, Patras,
  Greece\\ hskokos@pks.mpg.de}

\maketitle

\begin{abstract}
  We study the problem of efficient integration of variational
  equations in multi-dimensional Hamiltonian systems. For this
  purpose, we consider a Runge-Kutta-type integrator, a Taylor series
  expansion method and the so-called `Tangent Map' (TM) technique
  based on symplectic integration schemes, and apply them to the
  Fermi-Pasta-Ulam $\beta$ (FPU-$\beta$) lattice of $N$ nonlinearly
  coupled oscillators, with $N$ ranging from 4 to 20. The fast and
  accurate reproduction of well-known behaviors of the Generalized
  Alignment Index (GALI) chaos detection technique is used as an
  indicator for the efficiency of the tested integration
  schemes. Implementing the TM technique--which shows the best
  performance among the tested algorithms--and exploiting the
  advantages of the GALI method, we successfully trace the location of
  low-dimensional tori.
\end{abstract}


%
%
\section{Introduction}

From interactions of stars in galaxies to particle beams in high
energy accelerators, Hamiltonian mechanics is found at the very heart
of modeling and understanding dynamical processes. The necessity to
classify evermore complex systems in terms of stability and
predictability has lead to a wealth of methods discriminating chaotic
from regular behavior (see for example \cite[Sect.~7]{S10}). Most of
these techniques rely on the study of the time evolution of deviation
vectors of a given orbit to discriminate between the two regimes. The
time evolution of these vectors is governed by the so-called
variational equations.

In \cite{P1} the `Tangent Map' (TM) technique, an efficient and easy
to implement method for the integration of the variational equations
of Hamiltonian systems based on the use of symplectic integrators was
introduced.  In \cite{P1,P2} the TM method was applied mainly to
low-dimensional Hamiltonian systems of 2 and 3 degrees of freedom, and
proved to be very efficient and superior to other commonly used
numerical schemes, both with respect to its accuracy and its speed.

The scope of the present work is to extend these results by
investigating whether the efficiency of the TM method persists also
when multi-dimensional Hamiltonian systems are considered. The study
of such systems presents a challenging numerical task, which makes the
use of fast and accurate numerical tools imperative. In the present
paper we use as a toy model the famous Fermi-Pasta-Ulam $\beta$
(FPU-$\beta$) lattice, which is presented in Sect.~\ref{sec:FPU}. In
Sect.~\ref{sec:num_methods} the different numerical methods we use,
i.e.~the Generalized Alignment Index (GALI) chaos indicator, the TM
method and the SABA family of symplectic integrators, the Taylor
series integrator TIDES, and the general-purpose high-accuracy
Runge-Kutta integrator DOP853, are presented and their properties are
briefly discussed. Then, in Sect.~\ref{sec:efficiency} the numerical
results of the application of these numerical schemes for the
integration of variational equations of the FPU system are presented,
while in Sect.~\ref{sec:resultsFPU} the GALI method is used for
locating motion on low-dimensional tori. Finally, in
Sect.~\ref{sec:summary} we summarize our results.

%
%
\section{The FPU lattice}
\label{sec:FPU}

As a model of a multi-dimensional Hamiltonian system we consider the
FPU-$\beta$ lattice \cite{FPU,Ford92,Chaos50}, which describes a chain
of $N$ particles with nearest neighbor interaction. Regarding
numerical integration algorithms, the FPU lattice is a very
challenging problem, since it exhibits oscillations on largely
different time scales. The Hamiltonian of this $N$ degrees of freedom
($N$D) system as a function of the momenta $\bm p=(p_1,\ldots,p_N)$
and the coordinates $\bm q=(q_1,\ldots,q_N)$ is given by
\begin{equation}
  H_N = H_N(\bm p,\bm q) = \sum_{i=1}^{N} \frac{p_i^2}{2} + \sum_{i=0}^{N}
  \left[\frac{(q_{i+1}-q_i)^2}{2} + \frac{\beta (q_{i+1}-q_i)^4}{4} \right] \, .
\label{eq:FPUHam}
\end{equation}
In our study we impose fixed boundary conditions,
i.~e.~$q_0=q_{N+1}=0$, set $\beta=1.5$, and consider models whose
number of particles vary from $N=4$ up to $N=20$. We note that in
\cite{SBA08,P1} the particular case $N=8$ was studied in detail.

The Hamiltonian (\ref{eq:FPUHam}) can be split into two parts $A$ and
$B$, which respectively depend only on the momenta and the
coordinates, i.~e.~$H_N=A(\bm p)+B(\bm q)$.  The Hamilton's equations
of motion are
\begin{equation}
\label{eq:FPUeq}
\dot q_i = \frac{\partial H_N}{\partial p_i} = p_i \qquad \mathrm{and}\qquad \dot p_i=-\frac{\partial H_N}{\partial q_i}=(q_{j+1}-q_j)(\delta_j^i-\delta_{j+1}^i)+\beta(q_{j+1}-q_j)^3(\delta_j^i-\delta_{j+1}^i),
\end{equation}
with $1\leq i \leq N$, and $\delta_j^i$ denoting the Kronecker delta,
which is equal to 1 if $i=j$ and to 0 otherwise.  The variational
equations governing the time evolution of a deviation vector
$\vec{w}=(\delta q_1,\ldots,\delta q_N,\delta p_1,\ldots,\delta p_N)$,
that evolves in the tangent space of the Hamiltonian's phase space are
given by
\begin{equation}
\label{eq:FPUvareq}
\dot{\delta p_i} = \delta q_i \qquad \mathrm{and}\qquad \dot{\delta q_i} =-\sum_{j=1}^{N}  \frac{\partial^2 H_N}{\partial q_i \partial q_j}\delta q_j .
\end{equation}

%
%
\section{\label{sec:num_methods}Numerical methods}

In order to compute the stability of a particular solution of
Hamiltonian (\ref{eq:FPUHam}), or in other words of an orbit in the
2$N$-dimensional phase space of the system, the equations of motion
(\ref{eq:FPUeq}) have to be integrated together with the variational
equations (\ref{eq:FPUvareq}). The time evolution of the latter
contains information on the stability of the orbit, which can be
quantified by using some chaos indicator, for example the
GALIs. Different numerical approaches can be used to solve the system
of ordinary differential equations given by Eqs.~(\ref{eq:FPUeq}) and
(\ref{eq:FPUvareq}). In this section, after briefly recalling the
definition of the GALI and its behavior for regular and chaotic
motion, we present an overview of the methods used in the current
study. A more detailed description of further possibilities based on
symplectic methods can be found in \cite{P1}.

\subsection{\label{sec:gali}The Generalized Alignment Index (GALI)}

The GALI was originally introduced in \cite{GALI:2007} as an efficient
chaos detection method, generalizing a similar indicator called the
Smaller Alignment Index (SALI) \cite{S01,SkoAntoBouVrah,SABV04}. The
method has been applied successfully to different dynamical systems
for the discrimination between regular and chaotic motion, as well as
for the detection of regular motion on low dimensional tori
\cite{ChrisBou,SBA08,BouManChris,MR10,MA11,MSA11}.

For $N$D Hamiltonians the Generalized Alignment Index of order $k$
(GALI$_k$), $2 \leq k \leq 2N$, is determined through the evolution of
$k$ initially linearly independent deviation vectors $\vec{w}_k(0)$,
which are continually normalized, keeping their directions
intact. According to \citet{GALI:2007} GALI$_k$ is defined as the
volume of the $k$-parallelepiped having the $k$ unitary deviation
vectors $\hat{w}_i(t)=\vec{w}_i(t)/ \|\vec{w}_i(t) \|$,
$i=1,2,\ldots,k$, as edges.  GALI$_k$ is therefore determined through
the wedge product of these vectors
\begin{equation}
  \mbox{GALI}_k(t)=\| \hat{w}_1(t)\wedge \hat{w}_2(t)\wedge \cdots
  \wedge\hat{w}_k(t) \|,
\label{eq:GALI}
\end{equation}
with $\| \,\cdot \,\|$ denoting the usual norm.  The behavior of
GALI$_k$ for regular motion on an $s$-dimensional torus is given by
\cite{GALI:2007,ChrisBou,SBA08}
\begin{equation}
  \mbox{GALI}_k (t) \propto \left\{ \begin{array}{ll} \mbox{constant} & \mbox{if
        $2\leq k \leq s$} \\ \frac{1}{t^{k-s}} & \mbox{if $s< k \leq 2N-s$} \\
      \frac{1}{t^{2(k-N)}} & \mbox{if $2N-s< k \leq 2N$} \\
\end{array}\right. ,
\label{eq:GALI_order_all}
\end{equation}
while for chaotic orbits GALI$_{k}$ tends to zero
\textit{exponentially} following the law \cite{GALI:2007}
\begin{equation}
  \mbox{GALI}_k(t) \propto e^{-\left[ (\sigma_1-\sigma_2) + (\sigma_1-\sigma_3)+
      \cdots+ (\sigma_1-\sigma_k)\right]t},
\label{eq:GALI_chaos}
\end{equation}
where $\sigma_1, \ldots, \sigma_k$ are the first $k$ largest Lyapunov
characteristic exponents of the orbit.

We chose to apply the GALI method in order to check the accuracy of
the numerical integration of variational equations because this index
depends on the evolution of an ensemble of deviation vectors. Thus,
even small errors in the integration of these vectors are expected to
significantly influence the evolution of GALIs.  In particular, we
focus our attention on regular orbits lying on tori of various
dimensions $s$, for which the GALIs exhibit different evolutions
depending on $s$ and the order $k$ of the index. The possible
deviations of numerically evaluated GALIs from their expected
behaviors (\ref{eq:GALI_order_all}) will identify the inaccurate
integration of the deviation vectors far better than it would be the
case using chaotic orbits, for which all GALIs tend to zero
exponentially (\ref{eq:GALI_chaos}). Throughout the paper we denote
regular orbits on an $s$-dimensional torus of the $N$D system
(\ref{eq:FPUHam}) as $\mathcal T^N_s$, where $s$ can range from 2 to
$N$.

\subsection{\label{sec:tm}The Tangent Map method using symplectic
  algorithms}

Symplectic methods are often the preferred choice when integrating
dynamical problems, which can be described by Hamiltonian functions. A
thorough discussion of such methods can be found in
\citet{Hairer_etal_02}. Let us just mention some properties of
symplectic integrators which are of interest for our study. Symplectic
methods cannot be used with a trivial automated step size control.  As
a consequence, they are usually implemented with a fixed integration
step $\tau$.  Due to their special structure they preserve the
symplectic nature of Hamilton's equations intrinsically, which in turn
leads to results that are more robust for long integration times.  A
side-effect of structure preservation is that the error in energy
remains bounded irrespective of the total integration time.

In \cite{P1} it was shown that it is possible to integrate the
Hamilton's equations of motion and the corresponding variational
equations using the TM technique based on symplectic splitting
methods. Let us outline the basic idea behind the TM method, which is
founded on a general result stated for example in \cite{LR01}:
Symplectic integrators can be applied to systems of first order
differential equations $\dot {\bm X}=L \bm X$, that can be written in
the form $\dot {\bm X}=(L_A+L_B){\bm X}$, where $L,L_A,L_B$ are
differential operators defined as $L_\chi f=\{\chi,f\}$ and for which
the two systems $\dot {\bm X}=L_A {\bm X}$ and $\dot {\bm X}=L_B {\bm
  X}$ are integrable. Here $\{f,g\}$ are Poisson brackets of functions
$f(\bm q,\bm p)$, $g(\bm q,\bm p)$ defined as:
\begin{equation}
  \{ f,g\}=\sum_{l=1}^{N} \left( \frac{\partial f}{\partial p_l}
    \frac{\partial g}{\partial q_l} - \frac{\partial f}{\partial q_l}
    \frac{\partial g}{\partial p_l}\right). \label{eq:Poisson}
\end{equation}
The set of Eqs.~(\ref{eq:FPUeq}) is one example of such a system,
since the Hamiltonian (\ref{eq:FPUHam}) can be divided into two
integrable parts $A$ and $B$ with $H=A(\bm p)+B(\bm q)$ as already
noted. A symplectic splitting method separates the equations of motion
(\ref{eq:FPUeq}) into several parts, applying either the operator
$L_A$ or $L_B$. These are the equations of motion of the Hamiltonians
$A$ and $B$, which can be solved analytically, giving explicit
mappings over the time step $c_i\tau$, where the constants $c_i$ are
chosen to optimize the accuracy of the integrator. These mappings can
then be combined to approximate the solution for a time step
$\tau$. In \cite{P1} it was shown that the derivatives of these
mappings with respect to the coordinates and momenta of the system
(the so-called tangent maps) can be used to calculate the time
evolution of deviation vectors or, in other words, solve the
variational equations (\ref{eq:FPUvareq}).

In \cite{LR01} a family of symplectic splitting methods called
SABA$_n$ and SBAB$_n$ was introduced, with $n$ being the number of
applications of operators $L_A$ and $L_B$.  These integrators were
designed to have only positive intermediate steps. Since it is not
possible to construct symplectic integrators of order\footnote{In this
  work we call a symplectic integrator to be of order $n$, if it
  introduces an error of $\mathcal{O}(\tau^n)$ in the approximation of
  the real solution, with $\tau$ being the integration time step.}
$>2$ with this property \cite{suzuki1991}, small negative corrector
steps $C$ can be added before and after the main body of the
integrator to further increase the accuracy. In
Sect.~\ref{sec:efficiency} we test 3 different integrators, namely
SABA$_2$, SABA$_2$C and SBAB$_2$C, present the results of this
comparison and discuss the characteristics of these algorithms when
applied to the FPU system.

The fact, that standard adaptive stepping policy is not possible with
symplectic integration schemes necessitates an initial assessment of
stability for the algorithms used, in order to derive an upper bound
on the choice of step-size.  Following \citet{Hairer_etal_02} we note
that SBAB$_1$ is equivalent to the well known St\o rmer-Verlet or
leap-frog method, and SABA$_1$ to its adjoint.  In order to perform a
linear stability analysis of the FPU-$\beta$ lattice problem for
SABA$_2$, we introduce normal mode momenta $P_i$ and coordinates $Q_i$
as it was done for example in \cite{SBA08}. The unperturbed
($\beta=0$) Hamiltonian (\ref{eq:FPUHam}) can then be written as a sum
of the so-called \textit{harmonic energies} $E_i$, i.e.
\begin{equation}
  H=\sum_i^N E_i = \sum_i^N \frac{1}{2}(P^2_i+\omega_i^2Q^2_i), \qquad \omega_i=2sin\left(\frac{i\pi}{2(N+1)}\right), \label{harmen}
\end{equation}
where $\omega_i$ are the corresponding \textit{harmonic frequencies}.
With $M$ denoting the Jacobian of the numerical mapping from initial
coordinates and momenta to those at the next step for SABA$_2$ we have
  \begin{equation*}
\left[
\begin{array}{c}
  Q_i(\tau) \\
  P_i(\tau) \\
\end{array}
\right]=M \left[
\begin{array}{c}
  Q_i(0) \\
  P_i(0) \\
\end{array}
\right]
\end{equation*}
\begin{equation*}
M= \left[
\begin{array}{ccc}
  1-\tau ^2 \omega ^2(\frac{1}{2}+\frac{1}{4 \sqrt{3}})-
  \tau ^4 \omega ^4(\frac{1}{48}+\frac{1}{16 \sqrt{3}}) & \,\,\,\,\,\,\,\,&
  \tau -\frac{\tau ^3 \omega ^2}{8 \sqrt{3}} \\
  \tau  \omega ^2(-1+\frac{1}{2 \sqrt{3}})+
  \tau ^3 \omega ^4(\frac{1}{4}-\frac{1}{4 \sqrt{3}})+
  \tau ^5 \omega ^6(\frac{1}{48}-\frac{1}{24 \sqrt{3}})\quad & \,\,\,\, &
  1-\tau ^2 \omega ^2(\frac{1}{2}+\frac{1}{4 \sqrt{3}})-
  \tau ^4 \omega ^4(\frac{1}{48}+\frac{1}{16 \sqrt{3}})
\end{array}
\right].
\end{equation*}
The characteristic polynomial of matrix $M$ amounts to $\lambda
^2+\lambda \left(-2+\tau ^2 \omega ^2-\frac{\tau ^2 \omega ^2}{2
    \sqrt{3}}+\frac{\tau ^4 \omega ^4}{24}-\frac{\tau ^4 \omega ^4}{8
    \sqrt{3}}\right)+1$.  For the eigenvalues to be of modulus one,
given the harmonic frequencies (\ref{harmen}) satisfy $|\omega_i|\leq
2$, the maximum admissible step-size for SABA$_2$ will be
$\tau_\mathrm{max}\simeq 1.6$. Thus, in our study we always use $\tau
< \tau_\mathrm{max}$.

\subsection{\label{sec:TIDES}Taylor series methods-TIDES}

The basic idea of the so-called Taylor series methods is to
approximate the solution at time $t_i+\tau$ of a given $d$-dimensional
initial value problem
\begin{equation}
  \frac{\mathrm d \bm X(t)}{\mathrm d t} = \bm f(\bm X(t))\qquad\bm X\in \mathbb
  R^d,\, t\in \mathbb R
\end{equation}
from the $n^\mathrm{th}$ degree Taylor series of $\bm X(t)$ at
$t=t_i$:
\begin{equation}
\label{eq:taylor_series}
\bm X(t_i+\tau)\simeq \bm X(t_i)+ \tau\frac{\mathrm d\bm X(t_i)}{\mathrm
  dt}+\frac{\tau^2}{2!}\frac{\mathrm d^2\bm X(t_i)}{\mathrm dt^2}+\ldots+
\frac{\tau^n}{n!}\frac{\mathrm d^n\bm X(t_i)}{\mathrm dt^n}.
\end{equation}
For details see e.g.~\cite[Sect.~I.8]{Hairer_etal_93} and references
therein.  In this work we call an integrator being of order $n$, when
the first neglected term in this Taylor series expansion is of
$\mathcal O(\tau^{n+1})$. The computation of the derivatives can be
very cumbersome, depending on the structure of $\bm f$, and is done
efficiently using automatic differentiation techniques (see for
example \cite{barrio}).

In \cite{P2} two different publicly available implementations of the
Taylor method were used and compared regarding their reliability and
efficiency in the case of a $2$D Hamiltonian system. One of these
integrators, called TIDES\footnote{Freely available at
  \texttt{http://gme.unizar.es/software/tides}.}
\cite{barrio,abad_etal}, which showed better performance, will also be
used in this work as a representative of methods based on Taylor
series expansions.  TIDES comes as a \emph{Mathematica} notebook.
After inserting the differential equations one desires to integrate,
the notebook generates automatically all the necessary subroutines to
compute the given problem. The FORTRAN code produced by TIDES was
included into the existing testbed of our work without further
modification.

In order to obtain optimal results, the TIDES algorithm is free to
choose its order and step size during the whole integration
interval. We used one parameter only, the so-called one step accuracy
$\delta$, to control the numerical performance of the algorithm. To be
more precise, an integration step from time $t_i$ to $t_i+\tau$ is
accepted, if the internal accuracy checks estimate that the local
truncation error of the solution $\bm X(t_i+\tau)$ is less than
$\delta$. If this error is too large, the integrator automatically
tries to increase the internal order and/or adjust the step size
$\tau$. After each successful step the deviation vectors are
re-normalized.

Let us remark that another elegant way to express
Eq.~(\ref{eq:taylor_series}) can be achieved using the so called Lie
series formalism. Lie series have been rediscovered by
\cite{groebner1967} and used extensively in the field of dynamical
astronomy (e.g. \cite{hd1984,delva1984}) to numerically solve ordinary
differential equations. Sharing the same ansatz with symplectic maps
(Sect.~\ref{sec:tm}), Lie series can be used to iterate first order
ordinary differential equations of the form $\dot {\bm X}=L{\bm X}$ as
follows:
\begin{equation}
  {\bm X}_{t+\tau}= e^L {\bm X}_t = \sum_{j=0}^n \frac{(\tau L)^j}{j!} {\bm X}_t + \mathcal{O}(\tau^{n+1})
\end{equation}
Note that contrary to Sect.~\ref{sec:tm} no assumptions on the
properties of the differential operator are made. Thus, by evaluating
the consecutive derivatives $L{\bm X},\,L^2{\bm X},\,L^3{\bm X}$ and
building the corresponding exponential series up to order $n$ one is
able to follow the trajectory of ${\bm X}$ through phase space.  The
truncated Lie series' numerical map ${\bm X}_t \rightarrow {\bm
  X}_{t+\tau}$ is not area-preserving, in general, since the method is
non-symplectic. Therefore, the implementation of an adaptive step-size
into Lie series algorithms is possible without reservations. This can
be achieved via estimates on the size of the remainder of the
exponential series (see for example \cite{eggl2010}).

Algorithms using series expansions become most efficient when combined
with recursion relations, where higher order derivatives can be
calculated from previous ones. In the course of this work, also
extensive tests have been undertaken to adapt the Lie series method to
the FPU-$\beta$ lattice. Since for this problem such recursions are
not available, we used algebraic manipulation software to compute the
successive applications of the operator $L$ to ${\bm X}_t$, and
implemented the generated code into a FORTRAN program. Since this
approach proved to be reliable but computationally very expensive, we
do not include it in the discussion of our results in
Sect.~\ref{sec:results}.

\subsection{\label{sec:DOP853}General purpose integrators-DOP853}

In general, the computation of higher derivatives of functions $\bm
X(t)$ soon becomes very complicated. Therefore, the methods described
in Sect.~\ref{sec:TIDES} became popular only recently, after automatic
differentiation could be performed efficiently by computers. Before
that, other methods were developed to approximate
Eq.~(\ref{eq:taylor_series}). One of these are the so called
Runge-Kutta methods (see for example
\cite[Sect.~II.1.1]{Hairer_etal_93}). An $s$-stage Runge-Kutta method
is given as
\begin{eqnarray}
  \bm X(t_i+\tau) &=& \bm X(t_i) + \tau \sum_{i=1}^s b_ik_i\qquad\mathrm{with}\\ \nonumber
  k_i &=& \bm f(t_i+c_i\tau,\bm X(t_i)+\tau\sum_{j=1}^sa_{ij}k_j)\qquad\mathrm{and}\qquad c_i=\sum_{j=1}^s a_{ij}.
\end{eqnarray}
The real numbers $b_i,a_{ij}$ with $i,j=1,\ldots,s$ are chosen to
approximate Eq.~(\ref{eq:taylor_series}) to the desired order. If one
requires further $a_{ij}=0$ for $i\le j$ the integration method will
be explicit. Runge-Kutta integrators exist as symplectic as well as
non-symplectic variants.

In this work a 12-stage explicit Runge-Kutta integration method called
DOP853 is used\footnote{DOP853 is freely available from
  \texttt{http://www.unige.ch/\textasciitilde
    hairer/software.html}.}. This non-symplectic scheme is based on
the method of Dormand and Price (see \cite[Sect.~II.5]{Hairer_etal_93}
for further details). With this integrator we solve the set of
differential equations composed of Eqs.~(\ref{eq:FPUeq}) and
(\ref{eq:FPUvareq}) simultaneously. Here we use again the parameter
$\delta$ to control the integrator's overall behavior. For the DOP853
integrator the estimation of the local error and the step size control
is based on embedded formulas of orders 5 and 3.

%
%
\section{\label{sec:results}Results}

Before investigating how variational equations can be integrated
efficiently, one should first clarify the meaning of the term
`efficiency'. On the one hand, an efficient integration is one that is
performed as fast as possible, and on the other hand, this computation
should be also done as accurately as possible.  Since accuracy always
comes at the cost of intense computational efforts --meaning large CPU
times-- it tends to contradict the first mentioned aspect of
efficiency. In addition, since we are especially interested in the
computation of chaos indicators, an accurate computation should imply
the correct distinction between regular and chaotic orbits of the
studied dynamical system. Thus, in this work we consider the
integration of variational equations to be efficient, when the
computed GALIs are obtained with the least-possible CPU time
requirements, achieving at the same time the correct characterization
of orbits as regular or chaotic.

In the next section we present a thorough discussion of how
variational equations can be integrated numerically using the methods
described in Sect.~\ref{sec:num_methods}. We explain possibilities of
estimating the accuracy of such computations and discuss the obtained
results with respect to our definition of efficiency. In
Sect.~\ref{sec:resultsFPU} we apply this knowledge for a more global
investigation of the properties of the FPU-$\beta$ lattice. As a final
remark we note that all presented computations were performed on an
Intel Xeon X3470 with 2.93 GHz, using extended precision (80 bit).

\subsection{\label{sec:efficiency}Efficient integration of variational equations}

In most applications it is common to integrate the variational
equations together with the equations of motion they are based
on. Thus, an important prerequisite to obtain correct stability
results is the correct computation of the dynamics of the system
itself. For this reason we first discuss the properties of the
integrators described in Sect.~\ref{sec:num_methods} when applied to
some specific orbits of the FPU-$\beta$ lattice. In general, an
accuracy estimate of a numerically obtained orbit of a dynamical
system can be given via monitoring how well the system's first
integrals (e.g.~the total energy, total linear momentum etc.) are
conserved. For conservative Hamiltonian systems, like
(\ref{eq:FPUHam}), this can be done easily by checking the
conservation of the energy $H$ itself. The absolute value of the
relative errors of the total energy $|\Delta H/H|$, for different
integrators and step sizes $\tau$, for a $\mathcal T^4_2$ orbit over
$t=10^6$ time units are given in Table~\ref{tab1}. For non-symplectic
methods also the one-step accuracy $\delta$ is mentioned.

\begin{table} [htb]
  \tbl{\label{tab1}Information on the performance of the different numerical
    methods used for the computation of all the GALIs of the $\mathcal T^4_2$  orbit with initial condition  $q_i=0.1$, $p_i=0$, $1\leq i \leq 4$, of system (\ref{eq:FPUHam}) with $N=4$, over $t=10^6$ time units. 
    The given $\tau$ for non-symplectic methods is computed as a mean step size  $\tau=t/n_a$, where $n_a$ is the number of accepted steps, while $\delta$ is the one-step accuracy. 
    The absolute value of the relative energy error $|\Delta H/H|$ at the end of each integration, as well as the required CPU time for each method are also reported.
    The last column provides information on whether the computed GALIs identified the nature of the regular orbit correctly (Y)  or not (N) within the time interval $t=10^6$ (see also Fig.~\ref{fig1}).}
  {\begin{tabular}{llcclcc}
      Integrator & \multicolumn{1}{c}{$\delta$} & $\tau$ & Order & $|\Delta H/H|$ & CPU time & Correctness\\
      \toprule
      TM-SABA$_2$  &             & 1.00 &  2 & $6\times10^{-2}$  & 0 min 03 sec & Y\\
      TM-SABA$_2$  &             & 0.50 &  2 & $2\times10^{-3}$  & 0 min 06 sec & Y\\
      TM-SABA$_2$C &             & 0.50 &  4 & $4\times10^{-5}$  & 0 min 09 sec & Y \\
      TM-SABA$_2$C &             & 0.10 &  4 & $5\times10^{-7}$  & 0 min 46 sec & Y\\
      \\
      TIDES     & $10^{-5}$   & 0.66 & 10 & $4\times10^{-2}$  & 0 min 45 sec & N\\
      TIDES     & $10^{-8}$   & 0.60 & 14 & $1\times10^{-5}$  & 1 min 14 sec & N\\
      TIDES     & $10^{-10}$  & 0.54 & 16 & $2\times10^{-7}$  & 1 min 37 sec & Y\\
      \\
      DOP853    & $10^{-5}$   & 1.32 &  8 & $5\times10^{-1}$  & 0 min 11 sec & N\\
      DOP853    & $10^{-10}$  & 0.25 &  8 & $8\times10^{-6}$  & 0 min 54 sec & Y\\
      DOP853    & $10^{-11}$  & 0.19 &  8 & $6\times10^{-7}$  & 1 min 11 sec & Y\\
      \botrule
\end{tabular}}
\end{table}

Firstly, we focus on the application of the TM method with the SABA
and SBAB integrators.  The corresponding results are given in the
first 5 lines in Table~\ref{tab1}. Comparing the energy conservation
between SABA$_2$ and SABA$_2$C one finds that the use of the corrector
steps significantly improves $|\Delta H/H|$ for the same step size of
$\tau=0.5$. This result can of course be explained by the fact that
SABA$_2$ is an integration scheme of order 2, while SABA$_2$C is of
order 4. Reducing the step size for SABA$_2$C to $\tau=0.1$ further
improves the energy conservation as expected. We note that this
reduction leads to a linear growth by a factor 5 of the required CPU
time, as expected. Since SABA$_2$C shows the best performance, we use
only this integrator for further investigations.

Comparing the results of SABA$_2$C with the ones obtained by the
non-symplectic methods TIDES and DOP853 one finds that both
non-symplectic methods need more CPU time in order to reach the same
final value of $|\Delta H/H|$. If one computes a mean step size for
these algorithms, defined as the total integration time $t(=10^6)$
divided by the number of accepted steps $n_a$, one finds that both
integrators achieve this final energy error by a larger mean step
size, compared to SABA$_2$C, which is due to the higher orders of
these integrators. While the highest order for DOP853 is fixed to 8,
TIDES uses automatic order selection, which explains the larger mean
time step for the same value of one-step accuracy $\delta$.

While monitoring energy conservation serves as a control parameter
over the state vector of the system itself, it lacks information on
how accurately the corresponding variational equations are solved. If
the stability of certain initial conditions is known, one can use the
theoretically predicted behaviors (\ref{eq:GALI_order_all}) of the
GALI chaos indicator to estimate the reliability of the numerical
computation. Therefore, in the last column of Table~\ref{tab1}, we
provide information on whether the integration was able to identify
the $\mathcal T^4_2$ orbit correctly as being a regular orbit lying on
a 2-dimensional torus.

\begin{figure}[htb]
\begin{center}
\psfig{file=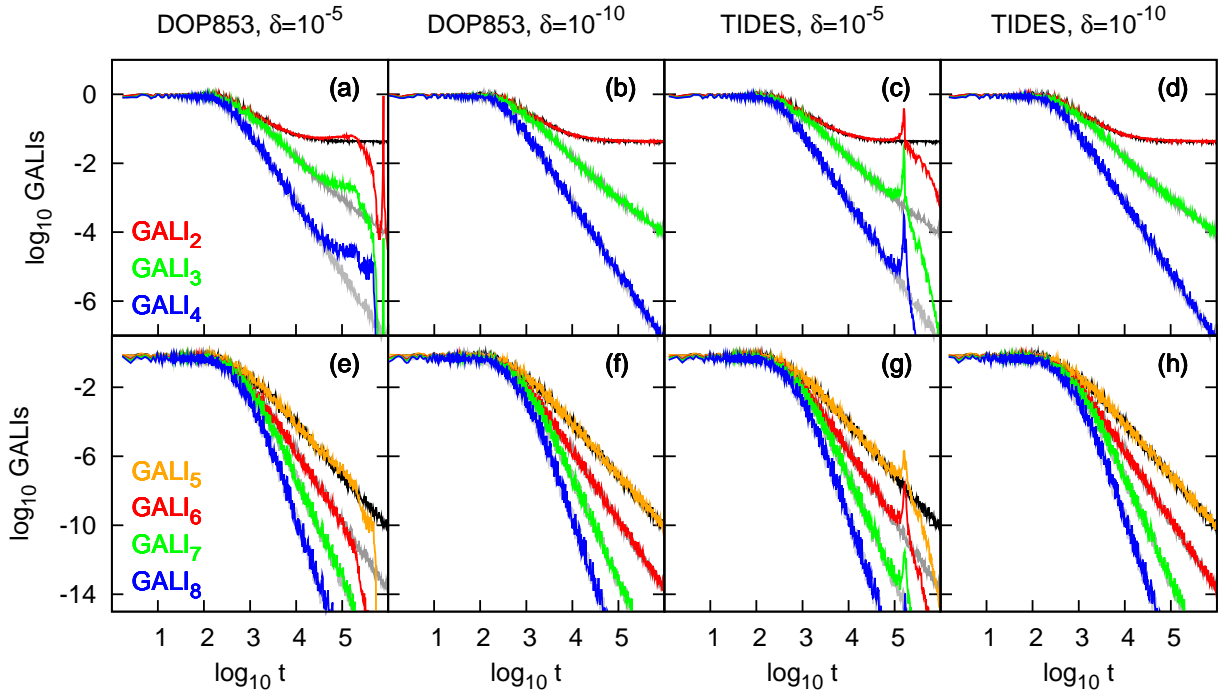,width=17cm} 
\end{center}
\caption{Time evolution of GALIs for the regular $\mathcal T^4_2$
  orbit with initial condition $q_i=0.1$, $p_i=0$, $1\leq i \leq 4$ of
  system (\ref{eq:FPUHam}) with $N=4$, as computed using
  non-symplectic schemes. The results are given as colored curves.
  The TM method results with SABA$_2$C and $\tau=0.5$ are presented by
  grey-scale curves in the background serving as a reference. These
  curves are not always clearly visible as they are overlapped by the
  colored ones.}
\label{fig1}
\end{figure}

The time evolution of GALI$_k$, $k=2,\ldots,8$, for some of the runs
of Table~\ref{tab1} is shown in Fig.~\ref{fig1}.  From
Eq.~(\ref{eq:GALI_order_all}) it is known that GALI$_2$ should be
constant for a $\mathcal T^4_2$ orbit, while GALI$_3$ and GALI$_4$
should decrease proportionally to $t^{-2}$ and $t^{-4}$
respectively. A correct characterization is possible by using the TM
method with SABA$_2$ and a step-size as large as $\tau=1.0$, although
the corresponding energy error $|\Delta H/H| \approx10^{-2}$, is
rather high (see first line in Table~\ref{tab1}). For the
non-symplectic methods it is found that $\delta=10^{-5}$ leads to a
similar error in energy conservation, but is not sufficient for the
correct computation of the GALIs (see the first and third columns of
Fig.~\ref{fig1}). Decreasing the one-step accuracy $\delta$ improves
the accuracy, but the integrations become less and less efficient
compared to the TM integration as the required CPU time grows.  We
note that a correct dynamical characterization of the orbit is
possible both for the DOP853 and TIDES for $\delta \lesssim 10^{-10}$.

\begin{figure}[htb]
\begin{center}
\psfig{file=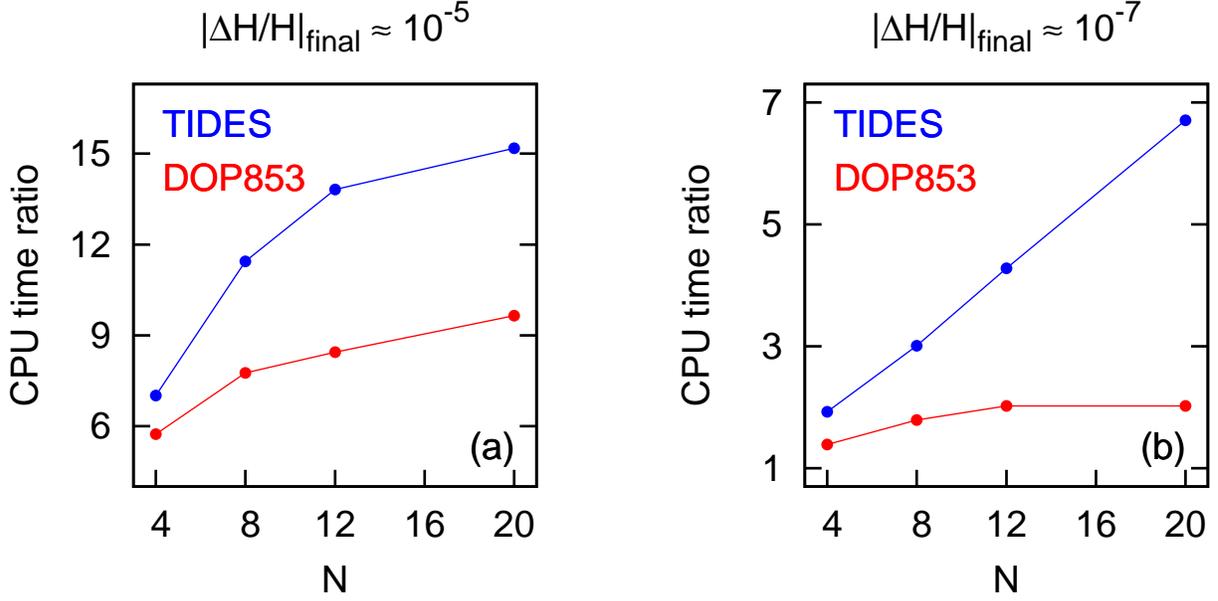,width=17cm} 
\end{center}
\caption{The ratio of required CPU time between integrations using the
  TM method with SABA$_2$C and the TIDES (blue curves) and DOP853 (red
  curves) integrators for a $\mathcal T^{N}_{N/2}$ regular orbit of
  Hamiltonian (\ref{eq:FPUHam}) with initial condition $q_i=0.1$,
  $p_i=0$, $1\leq i \leq N$ for $N=4$, 8, 12 and 20. Comparisons are
  performed between (a) TM method with step size $\tau=0.5$, TIDES
  with $\delta=10^{-8}$, and DOP853 with $\delta=10^{-10}$, and (b) TM
  method with $\tau=0.1$, TIDES with $\delta=10^{-10}$, and DOP853
  with $\delta=10^{-11}$.}
\label{fig2}
\end{figure}

Using energy conservation as an indicator for the quality of an
integration one could argue that the difference in CPU time between
SABA$_2$C with $\tau=0.1$ and DOP853 with $\delta=10^{-10}$ is not
very significant. While this is true for the $\mathcal T^4_2$ orbit,
the difference becomes more pronounced, when the number of particles
$N$ of the FPU-$\beta$ lattice is increased. This is evident in
Fig.~\ref{fig2} where we plot, as function of $N$, the ratio of the
required CPU times between the TM method with SABA$_2$C and the TIDES
(blue curves) and DOP853 (red curves) integrators for a $\mathcal
T^{N}_{N/2}$ regular orbit with initial condition $q_i=0.1$, $p_i=0$,
$1\leq i \leq N$ for $N=4$, 8, 12 and 20. If after $t=10^6$ time units
an error in energy conservation of $|\Delta H/H| \approx 10^{-5}$ is
sufficient, one could use either SABA$_2$C with $\tau=0.5$, TIDES with
$\delta=10^{-8}$ (although for $N=4$ the corresponding GALIs do not
exhibit the theoretically expected behavior for the whole time
interval), or DOP853 with $\delta=10^{-10}$ (see
Table~\ref{tab1}). The ratio of the CPU time of these runs is given in
Fig.~\ref{fig2}(a). In Fig.~\ref{fig2}(b) we show a similar comparison
when SABA$_2$C with $\tau=0.1$, TIDES with $\delta=10^{-10}$, and
DOP853 with $\delta=10^{-11}$ are used, which yields $|\Delta H/H|
\lesssim 10^{-6}$ at the end of the integration. In this case, the
GALIs computed by these methods show the time evolution predicted in
Eq.~(\ref{eq:GALI_order_all}).

Figure \ref{fig2} shows that the efficiency of the TM method improves
with increasing $N$ when compared to the non-symplectic methods used
in this study. While the ratio is 2 between the CPU times of SABA$_2$C
and TIDES for $N=4$ in Fig.~\ref{fig2}(b) it increases up to 7 when
using $N=20$ particles in the FPU-$\beta$ lattice. The ratios are
generally smaller when comparing the TM method with DOP853, but also
here a growing trend can be observed. The results become even more
pronounced when a larger error in energy conservation is acceptable,
as it is shown in Fig.~\ref{fig2}(a).

\begin{figure}[ht]
\begin{center}
\psfig{file=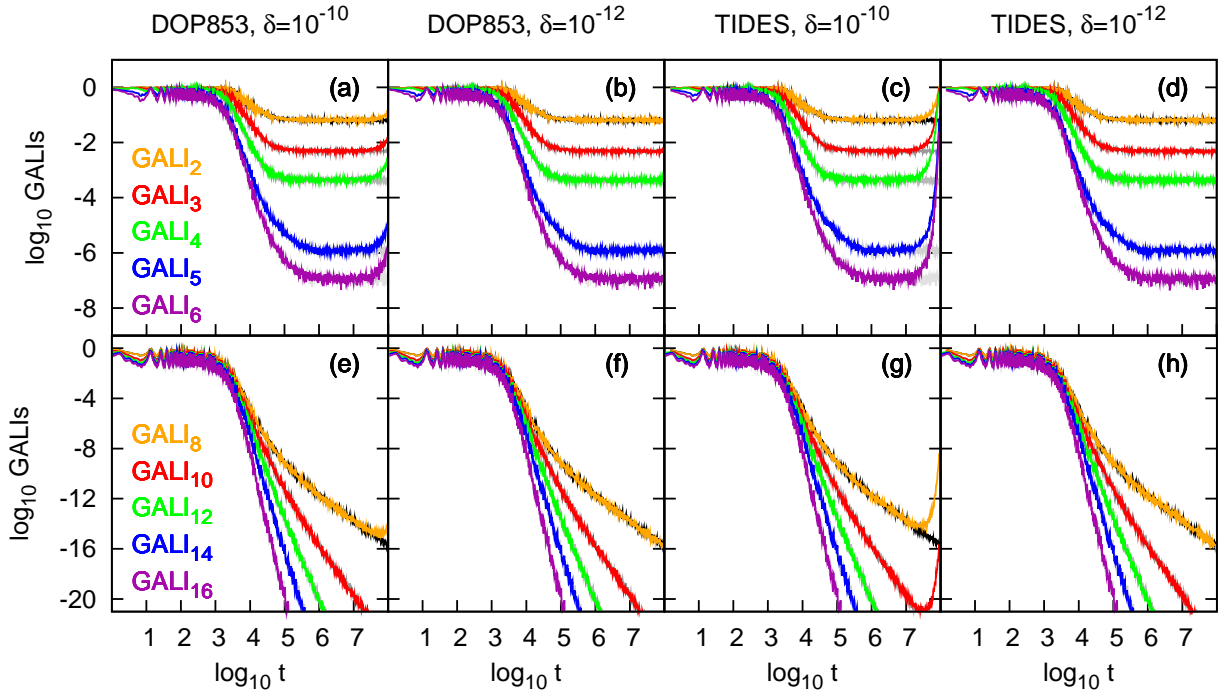,width=17cm} 
\end{center}
\caption{Time evolution of GALIs for a regular $\mathcal T^{12}_6$
  orbit with initial condition $q_i=0.1$, $p_i=0$, $1\leq i \leq 12$
  of system (\ref{eq:FPUHam}) with $N=12$, as computed using
  non-symplectic schemes. The results are given as colored curves,
  while the TM method results with SABA$_2$C and $\tau=0.1$ are given
  in grey-scale in the background as reference.}
\label{fig3}
\end{figure}

When analyzing dynamical systems one very often has to follow a
trajectory for long time intervals to determine the behavior of a
specific initial condition correctly. Especially for weakly chaotic
orbits differences to regular orbits are shown only in the late
evolution of chaos indicators. In such investigations the TM method
also proves to be superior compared to other techniques. As an example
we show in Fig.~\ref{fig3} the time evolution of GALIs for a regular
$\mathcal T^{12}_6$ orbit with initial condition $q_i=0.1$, $p_i=0$,
$1\leq i \leq 12$, for the system (\ref{eq:FPUHam}) with $N=12$, over
$t=10^8$ time units. Further information on these runs are given in
Table~\ref{tab2}.

\begin{table} [htb]
  \tbl{\label{tab2}Table similar to Table~\ref{tab1} containing information on the performance of the different numerical
    methods used for the computation of all the GALIs of the $\mathcal T^{12}_6$  orbit with initial condition  $q_i=0.1$, $p_i=0$, $1\leq i \leq 12$, of
    system (\ref{eq:FPUHam}) with $N=12$, over $t=10^8$ time units (see also Fig.~\ref{fig3}).}
  {\begin{tabular}{lcccccc}
      Integrator & $\delta$ & $\tau$ & Order & $|\Delta H/H|$ & CPU time & Correctness\\
      \toprule
      TM-SABA$_2$C &             & 0.50 &  4 & $3\times10^{-4}$  & 01 h 39 min 09 sec & Y \\
      TM-SABA$_2$C &             & 0.10 &  4 & $5\times10^{-7}$  & 08 h 02 min 17 sec & Y\\
      \\
      TIDES     & $10^{-10}$  & 0.54 & 16 & $1\times10^{-6}$  &  31 h 04 min 30 sec & N\\
      TIDES     & $10^{-12}$  & 0.51 & 22 & $1\times10^{-7}$  &  37 h 54 min 43 sec & Y\\
      \\
      DOP853    & $10^{-10}$  & 0.24 & 8 & $5\times10^{-4}$   & 12 h 45 min 07 sec & N\\
      DOP853    & $10^{-12}$  & 0.14 & 8 & $3\times10^{-6}$   & 22 h 30 min 19 sec & Y\\
      \botrule
\end{tabular}}
\end{table}

From Table~\ref{tab2} and Fig.~\ref{fig3} it is seen that for the
non-symplectic integrators a one-step accuracy of $\delta=10^{-10}$ is
not sufficient for a correct identification of the regular $\mathcal
T^{12}_6$ orbit. Results obtained both by the TIDES and the DOP853
integrators show a deviation from the theoretically predicted
behaviors (\ref{eq:GALI_order_all}) after $t=10^7$ (see the first and
third columns of Fig.~\ref{fig3}). To obtain the correct behavior of
GALIs over the whole integration interval it is necessary to decrease
the one-step accuracy to $\delta=10^{-12}$ (see the second and fourth
columns of Fig.~\ref{fig3}). We note that this decrease in $\delta$
naturally results in a significant increase in CPU time.  In contrast,
the GALIs obtained via the TM method with $\tau=0.5$ require
significantly less CPU time, and show the theoretically expected
behaviors up to $t=10^8$, although the relative energy error is
$|\Delta H/H|\approx10^{-4}$.

\begin{figure}[ht]
\begin{center}
\psfig{file=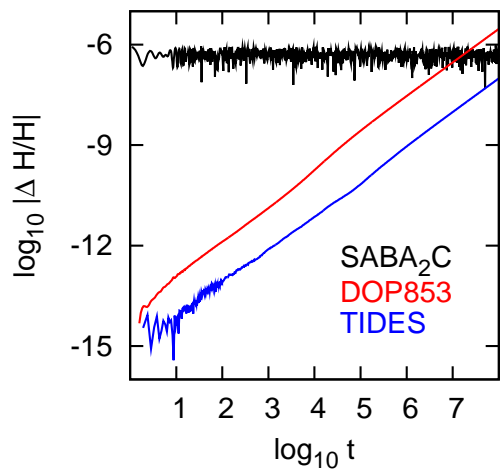,width=10cm} 
\end{center}
\caption{Time evolution of the absolute value of the relative error in
  energy conservation for the $\mathcal T^{12}_6$ orbit with initial
  condition $q_i=0.1$, $p_i=0$, $1\leq i \leq 12$, of Hamiltonian
  (\ref{eq:FPUHam}) with $N=12$. The symplectic routine SABA$_2$C uses
  a step size of $\tau=0.1$, while the non-symplectic methods required
  a one-step accuracy of $\delta=10^{-12}$ (see also
  Table~\ref{tab2}).}
\label{fig4}
\end{figure}

While the error in energy conservation $|\Delta H/H|$ grows with time
for non-symplectic integrators, it remains bounded for symplectic
integrators, as can exemplarily be seen in Fig.~\ref{fig4}. Thus, in
case it is necessary to integrate beyond $t=10^8$, one has to further
decrease $\delta$ for the non-symplectic methods, in order to achieve
the same final $|\Delta H/H|$. This, of course, would result in a
further increase in the ratio values of the CPU time required by the
non-symplectic methods as compared to the TM technique.

\subsection{\label{sec:resultsFPU}Searching for motion on
  low-dimensional tori}

One of the advantages of the GALI method is its capability to identify
motion on low-dimensional tori. For a regular orbit the largest order
$k$ of its GALIs that eventually remains constant determines the
dimension of the torus on which the motion occurs (see
Eq.~(\ref{eq:GALI_order_all})).  This ability was verified in
\cite{SBA08,BouManChris}, where some particular orbits on
low-dimensional tori were considered.

Since the TM method provides reliable evaluations of the GALIs even
for relatively large integration steps, and therefore requires little
CPU time, we applied this technique to perform a more global
investigation of the FPU-$\beta$ lattice, aiming to trace the location
of low-dimensional tori. In particular, we consider the Hamiltonian
system (\ref{eq:FPUHam}) with $N=4$, for which regular motion can
occur on an $s$-dimensional torus with $s=2$, 3, 4. According to
Eq.~(\ref{eq:GALI_order_all}) the corresponding GALIs of order $k \leq
s$ will be constant, while the remaining ones will tend to zero
following particular power laws. Thus, in order to locate
low-dimensional tori we compute the GALI$_k$, $k=2,3,4$ in the
subspace $(q_3,q_4)$ of the system's phase space, considering orbits
with initial conditions $q_1=q_2=0.1$, $p_1=p_2=p_3=0$, while $p_4$ is
computed to keep the total energy $H$ constant at $H=0.010075$.

\begin{figure}[htb]
\begin{center}
\psfig{file=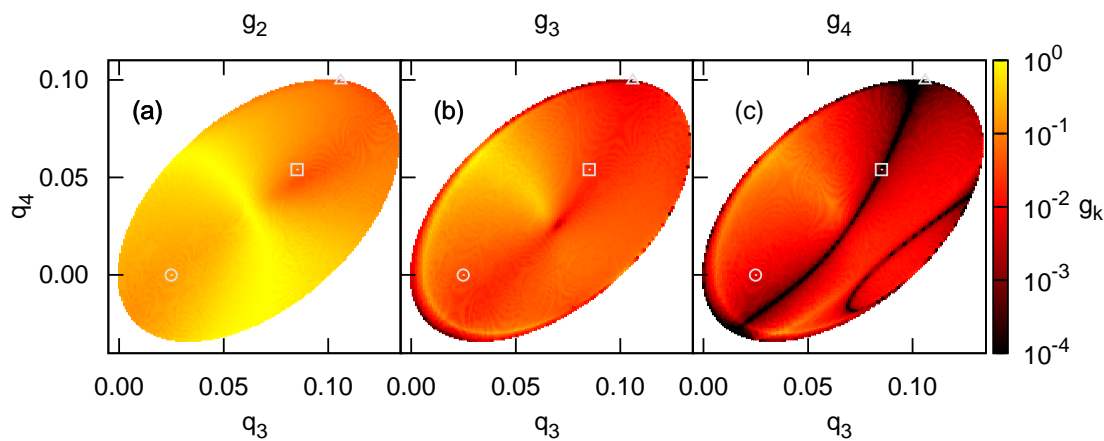,width=15cm} 
\end{center}
\caption{Regions of different $g_k$ (\ref{eq:GALIrescale}) values,
  $k=2$, 3, 4, on the $(q_3,q_4)$ plane of the Hamiltonian system
  (\ref{eq:FPUHam}) with $N=4$. Each initial condition is integrated
  by the TM method with SABA$_2$C and $\tau=0.5$ up to $t=10^6$, and
  colored according to its final (a) $g_2$, (b) $g_3$, and (c) $g_4$
  value, while white regions correspond to forbidden initial
  conditions. Three particular initial conditions of regular $\mathcal
  T^4_2$, $\mathcal T^4_3$ and $\mathcal T^4_4$ orbits are marked by a
  triangle, a square and a circle respectively.}
\label{fig5}
\end{figure}

Since the constant final values of GALI$_k$, $k=2,\ldots, s$, decrease
with increasing order $k$ (see for example the GALIs with $2 \leq k
\leq 6$ in Fig.~\ref{fig3}), we chose to `normalize' the values of
GALI$_k$, $k=2,3,4$ of each individual initial condition, by dividing
them by the largest GALI$_k$ value, $\mathrm{max}(\mathrm{GALI}_k)$,
obtained from all studied orbits. Thus, in Fig.~\ref{fig5} we colored
each initial condition according to its `\textit{normalized GALI$_k$}'
value
\begin{equation}
\label{eq:GALIrescale}
g_k = \frac{\mathrm{GALI}_k}{\mathrm{max}(\mathrm{GALI}_k)}.
\end{equation}

In each panel of Fig.~\ref{fig5}, large $g_k$ values (colored in
yellow or in light red) correspond to initial conditions whose
GALI$_k$ eventually stabilizes to constant, non-zero values. On the
other hand, darker regions correspond to small $g_k$ values, which
result from power law decays of GALIs.

\begin{figure}[htb]
\begin{center}
\psfig{file=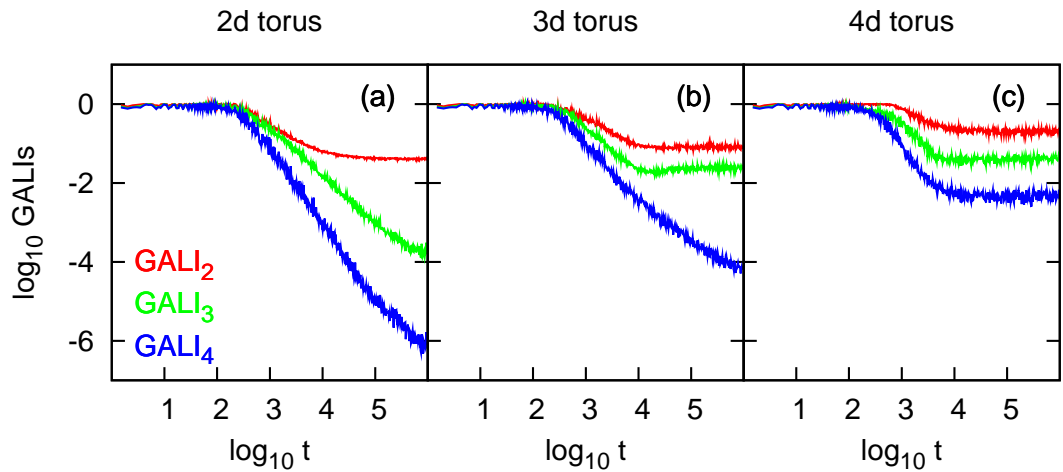,width=15cm} 
\end{center}
\caption{The time evolution of GALIs for regular orbits lying on a (a)
  2-dimensional torus, (b) 3-dimensional torus, and (c) 4-dimensional
  torus of the Hamiltonian system (\ref{eq:FPUHam}) with $N=4$. The
  initial conditions of these orbits are marked respectively by a
  triangle, a square and a circle in Fig.~\ref{fig5}.}
\label{fig6}
\end{figure}

Consequently, motion on 2-dimensional tori, which corresponds to large
final GALI$_2$ values and small final GALI$_3$ and GALI$_4$ values,
should be located in areas of the phase space colored in yellow or
light red in Fig.~\ref{fig5}(a), and in black in Figs.~\ref{fig5}(b)
and (c). A region of the phase space with these characteristics is for
example located in the upper border of the colored areas of
Fig.~\ref{fig5}. A particular initial condition with $q_3=0.106$,
$q_4=0.0996$ in this region is denoted by a triangle in
Fig.~\ref{fig5}. This orbit is indeed a $\mathcal T^4_2$ regular orbit
as we see from the evolution of its GALIs shown in
Fig.~\ref{fig6}(a). In a similar way, a $\mathcal T^4_3$ orbit should
be located in regions colored in black only in the $g_4$ plot of
Fig.~\ref{fig5}(c). An orbit of this type is the one with
$q_3=0.085109$, $q_4=0.054$ denoted by a square in Fig.~\ref{fig5},
which actually evolves on a 3-dimensional torus, as only its GALI$_2$
and GALI$_3$ remain constant (Fig.~\ref{fig6}(b)). Finally, the orbit
with $q_3=0.025$, $q_4=0$ (denoted by a circle in Fig.~\ref{fig5})
inside a region of the phase space colored in yellow or light red in
all panels of Fig.~\ref{fig5}, is a regular orbit on a 4-dimensional
torus and its GALI$_2$, GALI$_3$ and GALI$_4$ remain constant
(Fig.~\ref{fig6}(c)). It is worth noting, that chaotic motion would
lead to very small GALI$_k$ and $g_k$ values, since all GALIs would
tend to zero exponentially, and consequently would correspond to
regions colored in black in \textit{all} panels of
Fig.~\ref{fig5}. Thus, chaotic motion can be easily distinguished from
regular motion on low-dimensional tori.

From the results of Figs.~\ref{fig5} and \ref{fig6} it becomes evident
that the comparison of color plots of `normalized GALI$_k$' values can
facilitate the tracing of low-dimensional tori. The construction of
such plots becomes a very demanding computational task, especially for
high-dimensional systems. Thus, the application of the TM method for
obtaining such results becomes imperative, since the required
computations can be performed very efficiently by this method.

%
%
\section{Summary and Conclusions}
\label{sec:summary}

We compared different numerical techniques for the integration of
variational equations of multi-dimensional Hamiltonian systems. In
particular, we considered the TM method, which uses symplectic
integrators for the realization of this task, as well as
non-symplectic algorithms, like the general-purpose Runge-Kutta
integrator DOP853, and the TIDES algorithm, which relies on Taylor
series expansion techniques. These methods were applied to the $N$D
Hamiltonian system (\ref{eq:FPUHam}), the FPU-$\beta$ lattice, with
$N$ varying from $N=4$ to $N=20$.

We used the numerically obtained solutions of the variational
equations for the computation of the GALI chaos indicators. The
accurate reproduction of theoretically known behaviors of GALIs was
used as a measure of reliability of the numerical techniques
tested. In addition, the CPU time required by each method in order to
achieve accurate results, was taken into account for the
characterization of the efficiency of these algorithms.

The TM method exhibited the best numerical performance in all our
simulations, both in accuracy and speed. More specifically, we found
that the ratio of the CPU time required by the TIDES and DOP853
algorithms, with respect to the TM method, for correctly
characterizing the nature of orbits, increased with increasing $N$
(Fig.~\ref{fig2}). Thus, the TM method should be preferred over the
other two techniques, especially for studies of multi-dimensional
systems.

A feature of the TM method, which is of significant practical
importance, is that it succeeds in finding the correct GALI behavior,
and consequently determines the nature of orbits correctly, even when
large integration steps are used, despite the fact that in these cases
the energy accuracy is rather low. Therefore, the application of the
TM method allows the efficient investigation of the dynamical
properties of a large number of initial conditions in feasible CPU
times.  As an example, we showed in Sect.~\ref{sec:resultsFPU} how the
TM method can exploit the properties of GALI to efficiently find the
location of low dimensional tori in the phase space. Possible
applications of this approach could be the tracing of quasiperiodic
motion in multi-dimensional systems, when only a few degrees of
freedom are excited.

\nonumsection{Acknowledgments}

The work of E.~G.~was financially supported by the DFG research unit
FOR584.  S.~E.~would like to acknowledge the support of the Austrian
FWF project P20216.  Ch.~S.~was partly supported by the European
research project ``Complex Matter'', funded by the GSRT of the
Ministry Education of Greece under the ERA-Network Complexity Program.


\end{document}